\documentclass[twocolumn,english]{revtex4-1}
\usepackage[T1]{fontenc}
\usepackage[latin9]{inputenc}
\usepackage{amsmath}
\usepackage{graphicx}

\makeatletter
\usepackage{hyperref}

\makeatother

\usepackage{babel}
\begin{document}
\title{Data-driven criterion for the solid-liquid transition of two-dimensional
self-propelled colloidal particles far from equilibrium}
\author{Wei-chen Guo}
\affiliation{Guangdong Provincial Key Laboratory of Quantum Engineering and Quantum
Materials, School of Physics and Telecommunication Engineering, South
China Normal University, Guangzhou 510006, China}
\affiliation{Guangdong-Hong Kong Joint Laboratory of Quantum Matter, South China
Normal University, Guangzhou 510006, China}
\author{Bao-quan Ai}
\email{aibq@scnu.edu.cn}

\affiliation{Guangdong Provincial Key Laboratory of Quantum Engineering and Quantum
Materials, School of Physics and Telecommunication Engineering, South
China Normal University, Guangzhou 510006, China}
\affiliation{Guangdong-Hong Kong Joint Laboratory of Quantum Matter, South China
Normal University, Guangzhou 510006, China}
\author{Liang He}
\email{liang.he@scnu.edu.cn}

\affiliation{Guangdong Provincial Key Laboratory of Quantum Engineering and Quantum
Materials, School of Physics and Telecommunication Engineering, South
China Normal University, Guangzhou 510006, China}
\affiliation{Guangdong-Hong Kong Joint Laboratory of Quantum Matter, South China
Normal University, Guangzhou 510006, China}
\begin{abstract}
We establish an explicit data-driven criterion for identifying the
solid-liquid transition of two-dimensional self-propelled colloidal
particles in the far from equilibrium parameter regime, where the
transition points predicted by different conventional empirical criteria
for melting and freezing diverge. This is achieved by applying a hybrid
machine learning approach that combines unsupervised learning with
supervised learning to analyze a huge amount of the system's configurations
in the nonequilibrium parameter regime on an equal footing. Furthermore,
we establish a generic data-driven evaluation function, according
to which the performance of different empirical criteria can be systematically
evaluated and improved. In particular, by applying this evaluation
function, we identify a new nonequilibrium threshold value for the
long-time diffusion coefficient, based on which the predictions of
the corresponding empirical criterion are greatly improved in the
far from equilibrium parameter regime. These data-driven approaches
provide a generic tool for investigating phase transitions in complex
systems where conventional empirical ones face difficulties.
\end{abstract}
\maketitle

\section{Introduction}

For decades there have been open questions concerning the two-dimensional
(2D) solid-liquid transitions \citep{Strandburg_RMP_1988_2Dmelting}.
For instance, the well-known Kosterlitz-Thouless-Halperin-Nelson-Young
theory \citep{Kosterlitz_JPhysC_1973,Halperin_PRL_1978,Halperin_PRB_1979,Young_PRB_1979}
suggests that 2D crystals melt via two continuous transitions to liquid
with an intermediate hexatic phase, while first-order transitions
with and without the intermediate phase are also found both in numerical
simulations and in experiments in different 2D systems \citep{Krauth_PRL_2011,Krauth_PRL_2015,Russo_PRL_2017,Glotzer_PRX_2017,Guerra_Nature_2018,Digregorio_PRL_2018,Cugliandolo_PRL_2017,Komatsu_PRX_2015,Ning_Xu_PRL_2016,Ciamarra_PRL_2020,Krauth_Nat_Commun_2018,Caprini_PRR_2020}.

The situation becomes even trickier for 2D systems far from equilibrium.
For instance, the solid-liquid transitions of 2D nonequilibrium (NEQ)
colloidal crystals, ranging from 2D living crystals of photo-activated
colloidal surfers \citep{Palacci_Science_2013}, over schools of Janus
colloidal particles on a flat interface \citep{Dietrich_PRL_2018},
to active nematic liquid crystals of 2D epithelial tissues \citep{Saw_AM_2018},
etc., are typical representatives \citep{Lowen_PRL_2012}. In this
type of system, even though the transition points predicted by different
conventional empirical criteria \citep{Steinhardt_PRB_1983,Hartmann_PRE_2005,Lowen_PRL_1993,Lowen_PRE_1996,Zahn_PRL_2000}
for melting and freezing agree with each other in equilibrium, they
were shown to separate away when the system enters the NEQ parameter
regime, essentially resulting in a lower and an upper bound for the
solid phase region and the liquid one, respectively \citep{Lowen_PRL_2012}.
This thus raises the fundamental question of how to systematically
establish a criterion for identifying the solid-liquid transition
of 2D systems in NEQ and a way to evaluate the performance of different
empirical criteria.

In this work, we address this question for 2D self-propelled colloidal
particles with Yukawa-type interaction. To this end, we apply a hybrid
machine learning approach that combines unsupervised learning with
supervised learning to analyze a huge amount of the system's configurations
on an equal footing and establish an explicit data-driven criterion
for identifying the solid-liquid transition of the system in the far
from equilibrium parameter regime (cf.~Fig.~\ref{fig:main_results}).
More specifically, after generating $\sim\mathcal{O}(10^{6})$ spatial
distributions of colloidal particles in the NEQ steady state {[}cf.~the
first row of Fig.~\ref{fig:main_results}(a){]} via direct numerical
simulations of the dynamical equation (\ref{eq:Brownian_Yukawa})
of the system, these spatial distributions are analyzed by an unsupervised
learning approach \citep{van_Nieuwenburg_Nat_Phys_2017} that is realized
by a fully connected neural network (NN) \citep{Nielsen_Book_2015,Goodfellow_Book_2016}.
As a direct result, this gives an essentially unbiased (in the sense
of no prior empirical assumptions being involved) criterion to classify
the spatial distributions, according to which the solid-liquid transition
boundary is extracted {[}cf.~the solid curve in Fig.~\ref{fig:main_results}(b){]}.
Crucially, this also promotes the unlabeled data to the labeled data,
which enables the direct supervised learning to concretize the criterion
in the form of a set of NN-parameters that defines the NN {[}cf.~the
last row of Fig.~\ref{fig:main_results}(a){]}. This thus establishes
the explicit data-driven criterion that is able to identify the transition
not only within, but also beyond the NEQ parameter regime where the
supervised learning is directly performed {[}cf.~the stars in Fig.~\ref{fig:main_results}(b){]},
indicating that it indeed captures the essential feature of the solid-liquid
transition in the far from equilibrium parameter regime. Furthermore,
by utilizing the so-called classification accuracy generated in the
unsupervised learning process {[}cf.~Eq.(\ref{eq:Classification_accuracy})
and Fig.~\ref{fig:unsupervised_learning}(b){]}, we establish a generic
data-driven evaluation function {[}cf.~Eq.~(\ref{eq:criterial_classification_accuracy}){]},
according to which the performance of different empirical criteria
can be systematically evaluated and improved. The direct application
of this evaluation function gives rise, in particular, to a new NEQ
threshold value $D_{\mathrm{threshold}}^{\mathrm{NEQ}}$ for the long-time
diffusion coefficient, based on which the predictions of the corresponding
empirical criterion are greatly improved in the far from equilibrium
parameter regime {[}cf.~Fig.~\ref{fig:NEQ_D_threshold} and the
``$\times$'' marks in Fig.~\ref{fig:main_results}(b){]}. Moreover,
since the underlying machine learning techniques exploited in this
work are completely general, we expect that these data-driven approaches
can readily provide new physical insights into phase transitions in
other complex systems where conventional empirical approaches face
difficulties.

\section{Self-propelled colloidal particles and their 2D solid-liquid transition}

\begin{figure}
\noindent \begin{centering}
\includegraphics[width=3.3in]{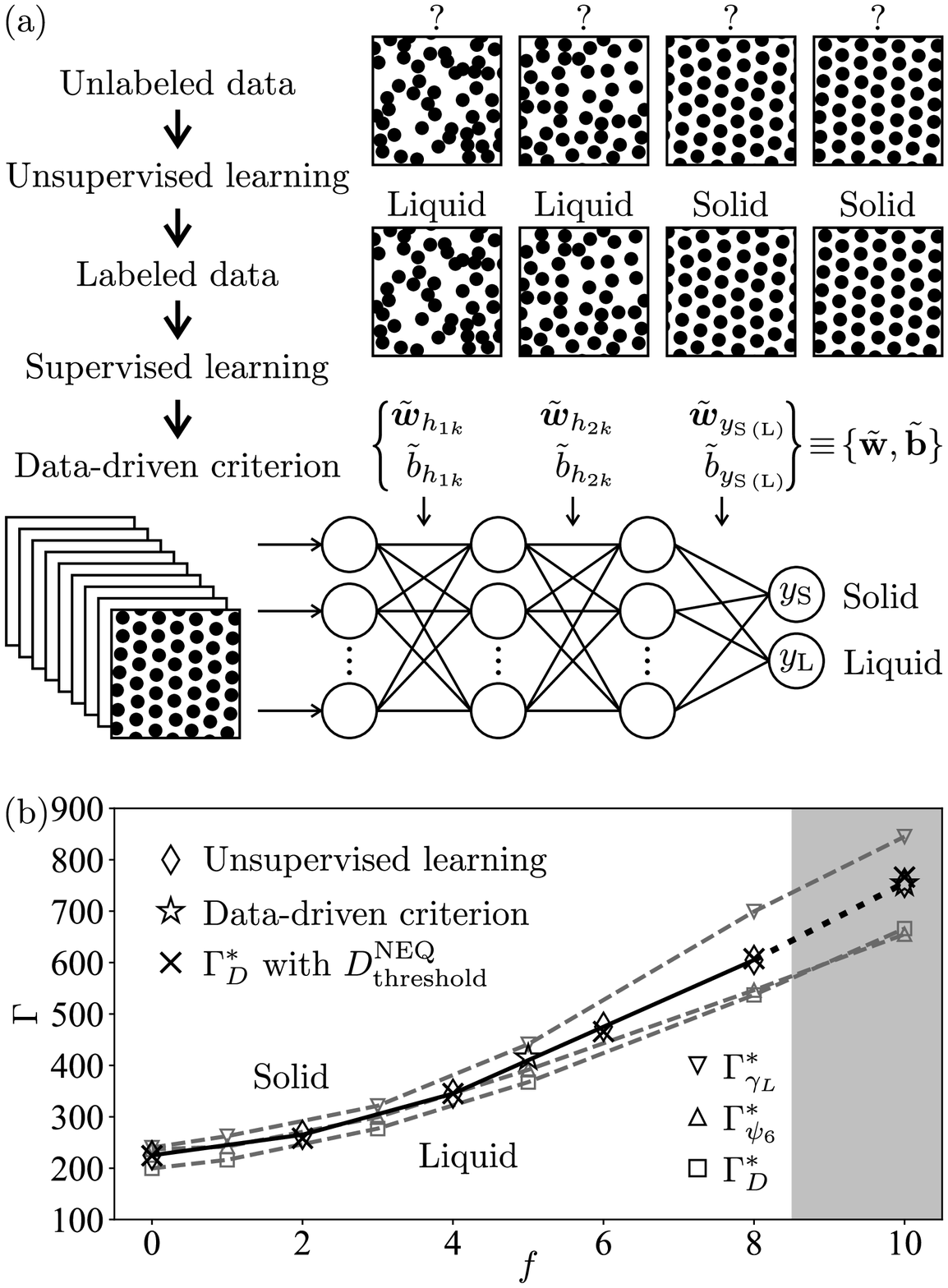}
\par\end{centering}
\caption{\label{fig:main_results}(a) Schematic illustration of the hybrid
machine learning approach. Spatial distributions of particles are
first processed by the unsupervised learning with a fully connected
NN, after which the unlabeled data are promoted to the labeled data.
These labeled data are processed by the supervised learning to optimize
the learnable parameter set of the NN and get the optimal set $\{\tilde{\mathbf{w}},\tilde{\mathbf{b}}\}$
that determines the data-driven criterion. (b) Solid-liquid transition
points predicted by the direct unsupervised learning (marked by ``$\diamondsuit$''
and solid curve), the data-driven criterion (marked by stars at $f=5$
and $f=10$), and the generalized empirical criterion with the NEQ
threshold $D_{\mathrm{threshold}}^{\mathrm{NEQ}}$ (marked by ``$\times$'').
For comparison, the transition points $\Gamma_{\psi_{6}}^{*}$, $\Gamma_{D}^{*}$,
and $\Gamma_{\gamma_{L}}^{*}$ predicted by the conventional empirical
criteria using $\psi_{6}$, $D$, and $\gamma_{L}$, respectively,
are shown by the dashed curves (reprinted from Ref.~\citep{Lowen_PRL_2012}).
The data-driven criterion and the NEQ threshold $D_{\mathrm{threshold}}^{\mathrm{NEQ}}$
are obtained by ``learning'' the configurations of the system with
$f=0,2,4,6,8$. Their predicted transition points match very well
with the ones from the direct unsupervised learning, not only in the
parameter regime $f\in[0,8]$ in which the data are directly accessible
to them, but also in the parameter regime well beyond (the shaded
area). See text for more details.}
\end{figure}

The system under study consists of $N$ self-propelled colloidal particles
in a 2D space \citep{Lowen_PRL_2012}, whose dynamics is described
by a set of overdamped Langevin-type equations with the explicit form,
\begin{equation}
\dot{\boldsymbol{r}}_{i}=-\nabla_{i}U+f\boldsymbol{e}_{i}+\boldsymbol{\xi}_{i}.\label{eq:Brownian_Yukawa}
\end{equation}
Here, $\boldsymbol{r}_{i}$ is the position of the $i$th particle,
and $\boldsymbol{\xi}_{i}$ is the noise which models the stochastic
interactions with the solvent molecules. The noise assumes zero mean
and correlations $\langle\boldsymbol{\xi}_{i}(t)\boldsymbol{\xi}_{j}^{T}(t^{\prime})\rangle=2\delta_{ij}\boldsymbol{1}\delta(t-t^{\prime})$,
where $\langle\cdot\rangle$ denotes an ensemble average over the
distribution of noise and $\delta(t)$ is the Dirac delta function.
$f$ is the strength of the force that propels each particle in the
direction $\boldsymbol{e}_{i}\equiv(\cos\varphi_{i},\sin\varphi_{i})$,
where $\langle\dot{\varphi}_{i}(t)\dot{\varphi}_{j}(t^{\prime})\rangle=2\Delta\delta_{ij}\delta(t-t^{\prime})$
with $\Delta$ being the so-called rotational diffusion coefficient.
$U=\sum_{i<j}u(\vert\boldsymbol{r}_{i}-\boldsymbol{r}_{j}\vert)$
is the interaction potential between particles, with $u(r)$ assuming
the form of the Yukawa potential, i.e., $u(r)=\Gamma\exp(-\lambda r)\slash r$.
Here, $\lambda$ is the inverse screening length, $\Gamma\equiv V_{0}\sqrt{\rho}\slash k_{B}T$
is the effective interaction strength, with $V_{0}$, $\rho$, and
$T$ being the bare potential strength, the density, and the temperature,
respectively.

By using empirical criteria associated with the global bond-orientational
order parameter $\psi_{6}$ \citep{Steinhardt_PRB_1983,Hartmann_PRE_2005},
the long-time diffusion coefficient $D$ \citep{Lowen_PRL_1993,Lowen_PRE_1996},
and the Lindemann-like parameter $\gamma_{L}$ \citep{Zahn_PRL_2000},
previous investigations have shown that for a fixed self-propelled
force strength $f$, the system assumes two phases, namely, the solid
phase at relatively large $\Gamma$ (corresponding to low temperature
or high density) and the liquid phase at relatively small $\Gamma$
(corresponding to high temperature or low density) \citep{Lowen_PRL_2012}.
However, it was also noted that the transition points predicted by
these different empirical criteria for melting and freezing, although
agree with each other in equilibrium ($f=0$), separate away when
the system enters the NEQ parameter regime ($f>0$), essentially resulting
in a lower and an upper bound for the solid phase region and the liquid
one \citep{footnote_1}, respectively {[}cf.~Ref.~\citep{Lowen_PRL_2012}
or the upper and lower dashed curves in Fig.~\ref{fig:main_results}(b){]}.

Noticing that the direct reliance on the properties of certain physical
quantities, namely, $\psi_{6}$, $D$, and $\gamma_{L}$, actually
makes the empirical criteria utilize only certain parts of the available
information of the system, it is intriguing to expect that a data-driven
criterion that directly makes full use of the available information
of the system would be able to identify the solid-liquid transition
in both equilibrium and NEQ. Indeed, as we shall see in the following,
such an explicit data-driven criterion can be established by directly
analyzing a huge amount of the system's configurations on an equal
footing. Moreover, a data-driven evaluation function can also be established,
according to which the performance of different empirical criteria
can be systematically evaluated and improved.

\section{Data-driven criterion for the solid-liquid transition via hybrid
machine learning approach}

\begin{figure}
\noindent \begin{centering}
\includegraphics[width=3.3in]{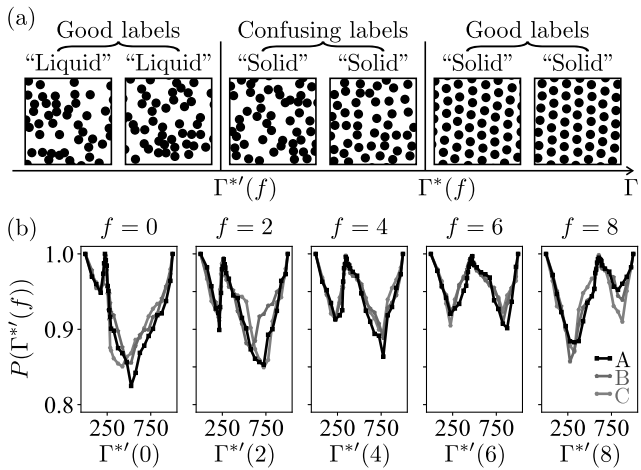}
\par\end{centering}
\caption{\label{fig:unsupervised_learning}(a) Schematic illustration of the
confusing labels involved in the unsupervised learning approach ``learning
by confusion''. The images with their corresponding $\Gamma$ satisfying
$\Gamma^{*\prime}(f)<\Gamma<\Gamma^{*}(f)$ are in the liquid phase
but labeled as ``solid'', which confuses the NN in establishing
the classification criterion. (b) Solid-liquid transition points identified
by the NN trained via unsupervised learning. For each self-propelled
force strength $f$, three independent data sets, denoted by ``A'',
``B'', and ``C'', are used in the unsupervised learning process,
and the classification accuracy $P(\Gamma^{*\prime}(f))$ for each
independent data set reaches the same nontrivial maximum that corresponds
to the solid-liquid transition point. The predicted transition points
are $\Gamma^{*}(0)=225\pm5$, $\Gamma^{*}(2)=265\pm5$, $\Gamma^{*}(4)=345\pm5$,
$\Gamma^{*}(6)=475\pm5$, and $\Gamma^{*}(8)=605\pm5$. See text for
more details.}
\end{figure}

The data-driven criterion is based on directly analyzing a huge amount
of the system's configurations, i.e., spatial distributions of particles,
which are generated by direct numerical simulations of the dynamical
equation (\ref{eq:Brownian_Yukawa}) of the system in different parameter
regimes. Here, we focus on the case with $N=1936$, $\Delta=3.5$,
$\lambda=3.5$, $\rho=1$, and all the numerical simulations of Eq.~(\ref{eq:Brownian_Yukawa})
are performed in a 2D rectangular space with aspect ratio $2/\sqrt{3}$
and periodic boundary condition imposed according to the minimum-image
convention \citep{Rapaport_art_of_MDS_CUP_2004}. In total, we generate
$\sim\mathcal{O}(10^{6})$ spatial distributions of self-propelled
colloidal particles in the steady state that correspond to different
sets of effective interaction strength $\Gamma$ and self-propelled
force strength $f$. These 2D distributions are then directly transformed
into images, forming the whole data set which is directly processed
by the NN that conducts the hybrid machine learning (cf.~Fig.~\ref{fig:main_results})
as we shall now discuss.

The hybrid machine learning starts with the unsupervised learning
performed by employing the so-called ``learning by confusion'' approach
\citep{van_Nieuwenburg_Nat_Phys_2017} that is realized by a fully
connected NN {[}cf.~Refs.~\citep{Nielsen_Book_2015,Goodfellow_Book_2016}
and the last row of Fig.~\ref{fig:main_results}(a){]}. Here, we
notice that although intermediate phases can appear from time to time
in various complex systems \citep{Glotzer_PRX_2017,Guerra_Nature_2018,Digregorio_PRL_2018,Cugliandolo_PRL_2017,Komatsu_PRX_2015,Ning_Xu_PRL_2016,Ciamarra_PRL_2020,Krauth_Nat_Commun_2018,Buttinoni_JPHYS_CM_2021},
for instance, in a related system with a relatively softer interaction
potential (inverse-power-law repulsion), the intermediate hexatic
phase was indeed identified in a quite narrow stripe region in the
phase diagram \citep{Krauth_Nat_Commun_2018}, previous investigations
have not resolve a possible intermediate phase for the finite-size
system under study in this work \citep{Lowen_PRL_2012} (see also
Appendix \ref{sec:scanning_probe}). Therefore, we use a relatively
simple binary classification approach, i.e., ``learning by confusion,''
to identify the possible solid-liquid transition, instead of resorting
to multiclass classification approaches, for instance, the one presented
in Appendix \ref{sec:scanning_probe}. This ``learning by confusion''
approach is an unsupervised learning approach in the sense that the
existence of possible transitions is not assumed \citep{van_Nieuwenburg_Nat_Phys_2017}
but determined by itself via monitoring the contrast between good
and bad recognition performance when ``confusing labels'' are deliberately
attached to the images in the data set (cf.~Fig.~\ref{fig:unsupervised_learning}),
despite being limited to investigate the cases where maximally two
possible phases exist in the parameter regimes of interests. Moreover,
since it is based on NN, a wide range of well-established powerful
NN architectures can be employed in its realization (see Appendix~\ref{sec:Convolutional-neural-network}
for a realization using a convolutional NN).

To implement the ``learning by confusion'' approach, images are
labeled as ``liquid'' (``solid'') if their corresponding $\Gamma<\Gamma^{*\prime}(f)$
($\Gamma>\Gamma^{*\prime}(f)$) by imposing a testing binary classification
rule associated with a proposed critical value $\Gamma^{*\prime}(f)$
for the images corresponding to the same fixed self-propelled force
strength $f$. This way of labeling generally confuses the NN when
the proposed critical value $\Gamma^{*\prime}(f)$ is not equal to
the physical critical value $\Gamma^{*}(f)$. For instance, if $\Gamma^{*\prime}(f)<\Gamma^{*}(f)$,
the images with their corresponding $\Gamma$ satisfying $\Gamma^{*\prime}(f)<\Gamma<\Gamma^{*}(f)$
are in the liquid phase but labeled as ``solid'' {[}cf.~Fig.~\ref{fig:unsupervised_learning}(a){]}.
For a generic proposed value $\Gamma^{*\prime}(f)$, we train the
NN and then test its classification accuracy,
\begin{equation}
P(\Gamma^{*\prime}(f))\equiv\frac{\mathcal{M}_{f}^{\textrm{``correct''}}(\Gamma^{*\prime}(f))}{\mathcal{M}_{f}^{\textrm{test}}},\label{eq:Classification_accuracy}
\end{equation}
with new testing images that have not been ``seen'' by the NN in
the training process (see Appendix \ref{sec:Training-process-of-NN}
for details). Here, $\mathcal{M}_{f}^{\textrm{test}}$ is the total
number of the testing images and $\mathcal{M}_{f}^{\textrm{``correct''}}(\Gamma^{*\prime}(f))$
is the number of images that are classified ``correctly'' according
to the proposed labels determined by $\Gamma^{*\prime}(f)$. Noticing
that the closer $\Gamma^{*\prime}(f)$ is to the solid-liquid transition
point $\Gamma^{*}(f)$, the fewer confusing labels exist, hence leading
to the relatively higher accuracy, the classification accuracy $P(\Gamma^{*\prime}(f))$
should therefore assume a nontrivial maximum when $\Gamma^{*\prime}(f)$
matches $\Gamma^{*}(f)$.

As we can see from Fig.~\ref{fig:unsupervised_learning}(b), where
the unsupervised learning is performed separately for the data sets
with corresponding $f=0,2,4,6,8$, the classification accuracy $P(\Gamma^{*\prime}(f))$
assumes a nontrivial maximum for each $f$, from which one can directly
read out the corresponding transition point $\Gamma^{*}(f)$. The
solid-liquid transition boundary formed by these transition points
is shown in the phase diagram Fig.~\ref{fig:main_results}(b), where
we can see that it locates exactly between the lower bound of the
solid region set by $\Gamma_{\gamma_{L}}^{*}$ and the upper bound
of the liquid region set by $\Gamma_{D}^{*}$, clearly corroborating
the predictions from these conventional dynamical criteria \citep{Lowen_PRL_2012}
and suggesting there exists an underlying criterion that can capture
the feature for distinguishing the solid from the liquid phase in
the far from equilibrium parameter regime. This thus motivates us
to extract the explicit form of this underlying criterion revealed
by the unsupervised learning.

To achieve this goal, we first notice that the explicit form of the
criterion can be generally regarded as a map from the space of the
system's configurations to two ``confidence'' values, say, $y_{\textrm{S}}$
and $y_{\textrm{L}}$ that hold the meaning of how likely a given
configuration is a configuration of the solid and the liquid phase,
respectively, i.e., a well-trained NN that takes the system's configurations
as input and outputs the correct confidence values {[}cf.~the last
row of Fig.~\ref{fig:main_results}(a){]}. Crucially, according to
the solid-liquid transition boundary predicted by the unsupervised
learning, all available images with corresponding $f=0,2,4,6,8$,
can be properly labeled now, thus promoting the original unlabeled
data set to the labeled one. With the labeled data, the supervised
learning realized by a fully connected NN with the learnable parameter
set $\left\{ \mathbf{w},\mathbf{b}\right\} $ can be directly performed
and finally gives rise to the optimal learnable parameter set $\{\tilde{\mathbf{w}},\tilde{\mathbf{b}}\}$
that completely determines the well-trained NN (see Appendix \ref{sec:NNs-confidence-values}
for details), i.e., the explicit data-driven criterion.

From Fig.~\ref{fig:main_results}(b) we can see that the prediction
on the solid-liquid transition points from this data-driven criterion
(see Appendix \ref{sec:Transition-points-predicted} for details)
matches very well with the one from the direct unsupervised learning.
In particular, for the transition point at $f=10$, it is located
way beyond the parameter regime $f\in[0,8]$ in which the data are
directly provided to the NN to be learned, clearly manifesting that
this explicit data-driven criterion indeed captures the generic feature
that distinguishes the solid from the liquid phase in the far from
equilibrium parameter regime. Moreover, noticing that essentially
only a lower bound for the solid phase region and an upper bound for
the liquid phase region were identified by the conventional empirical
approaches \citep{Lowen_PRL_2012}, the data-driven approach thus
establishes the solid-liquid transition boundary in both the equilibrium
and NEQ regimes for this system for the first time.

\section{Data-driven evaluation function for empirical criteria}

\begin{figure}
\noindent \begin{centering}
\includegraphics[width=3.3in]{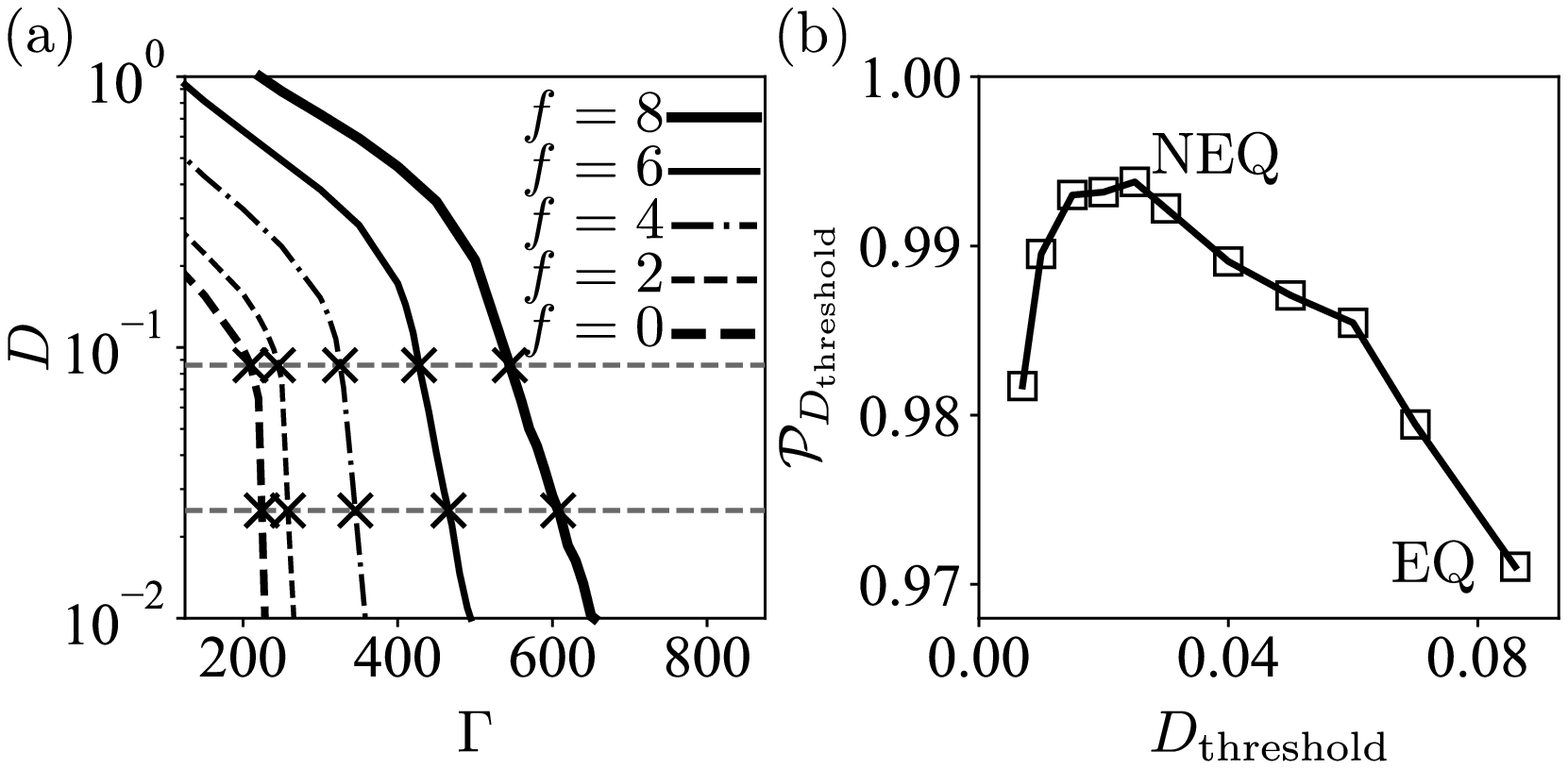}
\par\end{centering}
\caption{\label{fig:NEQ_D_threshold}(a) $\Gamma$-dependence of the long-time
diffusion coefficient $D$ of the system at different self-propelled
force strengths $f=0,2,4,6,8$. The upper and the lower horizontal
dashed lines correspond to the equilibrium threshold $D_{\mathrm{threshold}}^{\mathrm{EQ}}=0.086$
and the NEQ one $D_{\mathrm{threshold}}^{\mathrm{NEQ}}=0.025$ obtained
by optimizing $\mathcal{P}_{D_{\mathrm{threshold}}}$, respectively.
The crossings marked by ``$\times$'' of each horizontal dashed
line with the $D(\Gamma)$ curves at different $f$ predict the corresponding
transition points $\Gamma_{D_{\mathrm{threshold}}}^{*}(f)$. (b) Threshold
value $D_{\mathrm{threshold}}$ dependence of the criterial classification
accuracy $\mathcal{P}_{D_{\mathrm{threshold}}}$. The rightmost point
corresponds to $D_{\mathrm{threshold}}=D_{\mathrm{threshold}}^{\mathrm{EQ}}=0.086$,
and the point with the highest $\mathcal{P}_{D_{\mathrm{threshold}}}$
corresponds to $D_{\mathrm{threshold}}^{\mathrm{NEQ}}=0.025$ which
results in a generalized empirical criterion with improved performance
in the NEQ scenario. See text for more details.}
\end{figure}

Besides the solid-liquid transition boundary in the far from equilibrium
regime and its associated data-driven criterion, another class of
physical insights that one can obtain by making full use of the available
information of the system is a data-driven evaluation function, according
to which the performance of different empirical criteria can be systematically
evaluated and improved. From Eq.~(\ref{eq:Classification_accuracy}),
we can see that for any given (empirical) criterion, denoted as $\mathcal{C}$,
its performance can be naturally evaluated by the average classification
accuracy of its predicted transition points at which a sufficient
amount of the system's configurations are available, i.e., a criterial
classification accuracy denoted as $\mathcal{P}_{\mathcal{C}}$ with
the explicit form,

\begin{equation}
\mathcal{P}_{\mathcal{C}}=\frac{1}{N_{f}}\sum_{i=1}^{N_{f}}P(\Gamma_{\mathcal{C}}^{*}(f_{i})),\label{eq:criterial_classification_accuracy}
\end{equation}
where $\Gamma_{\mathcal{C}}^{*}(f_{i})$ is the transition point determined
by $\mathcal{C}$ at the self-propelled force strength $f_{i}$ and
$N_{f}$ is the total number of the transition points involved in
the evaluation.

Taking the empirical criterion that uses an equilibrium threshold
value $D_{\mathrm{threshold}}^{\mathrm{EQ}}=0.086$ for the long-time
diffusion coefficient $D$ \citep{Lowen_PRL_2012,Lowen_PRE_1996},
for example, by utilizing the available data at $f=0,2,4,6,8$, its
criterial classification accuracy $\mathcal{P}_{D_{\mathrm{threshold}}^{\mathrm{EQ}}}=\sum_{f=0,2,4,6,8}P(\Gamma_{D_{\mathrm{threshold}}^{\mathrm{EQ}}}^{*}(f))/5$
can be directly calculated, where $\Gamma_{D_{\mathrm{threshold}}^{\mathrm{EQ}}}^{*}(f)$
is the transition point at $f$ determined by this criterion {[}cf.~upper
row of ``$\times$'' in Fig.~\ref{fig:NEQ_D_threshold}(a) and
the rightmost point in Fig.~\ref{fig:NEQ_D_threshold}(b){]}. Noticing
this criterion employs a threshold value motivated by the equilibrium
behavior \citep{Lowen_PRL_2012,Lowen_PRL_1993,Lowen_PRE_1996}, hence
leaves the space for further improvement if a general threshold value
$D_{\mathrm{threshold}}$ is chosen {[}cf.~the lower row of ``$\times$''
in Fig.~\ref{fig:NEQ_D_threshold}(a){]}, we investigate the $D_{\mathrm{threshold}}$
dependence of the corresponding criterial classification accuracy
$\mathcal{P}_{D_{\mathrm{threshold}}}$. From Fig.~\ref{fig:NEQ_D_threshold}(b)
we notice that at a smaller threshold than the equilibrium one, $\mathcal{P}_{D_{\mathrm{threshold}}}$
reaches the highest value, directly indicating that the corresponding
criterion using this NEQ threshold, denoted as $D_{\mathrm{threshold}}^{\mathrm{NEQ}}$,
assumes the best performance in the NEQ parameter regime $f\in[0,8]$.
Indeed, the transition points predicted by this generalized empirical
criterion are surprisingly close to the ones obtained by the direct
unsupervised learning {[}cf.~Fig.~\ref{fig:main_results}(b){]}.
More remarkably, although $D_{\mathrm{threshold}}^{\mathrm{NEQ}}$
is determined within the NEQ parameter regime $f\in[0,8]$, its predictive
power extends well beyond, as manifested by its prediction of the
transition point at $f=10$ which is still very close to the one via
the direct unsupervised learning {[}cf.~the shaded area in Fig.~\ref{fig:main_results}(b){]}.
This thus suggests that after generalizing this criterion \citep{Lowen_PRL_2012,Lowen_PRL_1993,Lowen_PRE_1996}
with a smaller NEQ threshold, it can identify the solid-liquid transition
in both equilibrium and NEQ quite precisely, which is a clear improvement
over the conventional empirical criterion, made by utilizing the data-driven
evaluation function developed in this work. In fact, the improvement
is also consistent with the natural expectation that, due to the fact
that active systems in NEQ are more difficult to crystallize, the
threshold value for the general NEQ scenario is supposed to be smaller
than the one for the equilibrium case.

Moreover, we also perform a similar investigation on the empirical
criterion based on the threshold value of the global bond-orientational
order parameter $\psi_{6}$ \citep{Lowen_PRL_2012} (see Appendix
\ref{sec:Criterial-classification-accuracy} for details) and find
this criterion assumes relatively poor performance in general. Noticing
that being a static structure criterion by construction as well, the
data-driven criterion can indeed extract the solid-liquid transition
boundary in both equilibrium and NEQ, this thus manifests that it
is possible to develop a better static structure order parameter than
$\psi_{6}$. Comparing the data-driven criterion and the global bond-orientational
order parameter $\psi_{6}$, an obvious difference is that the former
directly makes full use of the available information of the system
while the latter essentially only utilizes information from groups
consisting of the six nearest neighboring particles \citep{Lowen_PRL_2012,Steinhardt_PRB_1983,Hartmann_PRE_2005}.
Therefore, one may expect that a more delicate static structure criterion
with, for instance, structural defects such as dislocations, disclinations,
etc., taken into account could improve the performance. Although constructing
a better structural order parameter is beyond the scope of the current
work, its performance can still be straightforwardly evaluated via
the criterial classification accuracy $\mathcal{P}_{\mathcal{C}}$
established in this work. More generally, the data-driven evaluation
function in fact offers a new type of generic tool to access crucial
insights into the performance of existing (empirical) criteria and,
in particular, facilitates the search for better criteria or order
parameters associated with various phase transitions in complex systems.

\section{Conclusion and outlook}

By directly analyzing over one million of the system's configurations
on an equal footing via data-driven approaches, new physical insights
into the solid-liquid transition of the system in the far from equilibrium
parameter regime are gained: An explicit data-driven criterion together
with its predicted solid-liquid transition boundary is established
via the hybrid machine learning approach that combines unsupervised
learning with supervised learning. Furthermore, the data-driven criterial
classification accuracy is established as a systematic way to evaluate
and improve the empirical criteria, via which, in particular, a generalized
conventional empirical criterion with a new NEQ long-time diffusion
coefficient threshold is found and assumes a much enhanced performance
in NEQ. These data-driven approaches open up a wide range of intriguing
possibilities for further investigations to provide new physical insights
into phase transitions in complex systems such as amorphous glass-formers
\citep{Biroli_Nat_Phys_2008,Albert_Science_2016} where conventional
empirical approaches face difficulties.
\begin{acknowledgments}
We thank Danbo Zhang and Fujun Lin for useful discussions. This work
was supported by NSFC (Grants No.~11874017 and No.~12075090), GDSTC
(Grants No.~2018A030313853 and No.~2017A030313029), GDUPS (2016),
Major Basic Research Project of Guangdong Province (Grant No.~2017KZDXM024),
Science and Technology Program of Guangzhou (Grant No.~2019050001),
and START grant of South China Normal University.
\end{acknowledgments}

\appendix

\section{NN's CONFIDENCE VALUES DETERMINED BY THE LEARNABLE PARAMETER SET\label{sec:NNs-confidence-values}}

In this work, each sample of the data set is an image of $K_{0}=3\times224\times224$
pixels which can be directly analyzed by a wide range of standard
NNs. In practice, this makes the test of robustness of the physical
results obtained by the machine learning approach in this work against
different NN architectures relatively convenient. Here, we choose
the employed fully connected NN \citep{Nielsen_Book_2015,Goodfellow_Book_2016}
to consist of an input layer $\boldsymbol{x}\equiv(x_{1},x_{2},\cdots x_{k},\cdots,x_{K_{0}})^{T}$
with $K_{0}$ neurons, two fully connected hidden layers $\boldsymbol{h}_{1}\equiv(h_{11},h_{12},\cdots h_{1k},\cdots,h_{1K_{1}})^{T}$
and $\boldsymbol{h}_{2}\equiv(h_{21},h_{22},\cdots,h_{2k},\cdots h_{2K_{2}})^{T}$
with $K_{1}=K_{2}=2\times10^{3}$ neurons, and an output layer with
$2$ neurons $y_{\textrm{S}}$ and $y_{\textrm{L}}$. When a sample
of our data, i.e., an image $I$ of the self-propelled colloidal particles'
spatial distribution, is fed to the fully connected NN, each neuron
$x_{k}$ in the input layer $\boldsymbol{x}$ collects a single raw
pixel of $I$ and delivers its value to the next layer. Each neuron
$h_{1k}$ in the next layer, namely, the first hidden layer $\boldsymbol{h}_{1}$,
receives $\boldsymbol{x}$ and takes the value $h_{1k}=\textrm{ReLU}(\boldsymbol{w}_{h_{1k}}^{T}\boldsymbol{x}+b_{h_{1k}})\equiv\max(0,\boldsymbol{w}_{h_{1k}}^{T}\boldsymbol{x}+b_{h_{1k}})$,
where the rectified linear unit (ReLU) is a nonlinear activation function
\citep{Nielsen_Book_2015,Goodfellow_Book_2016}. Similarly, each neuron
$h_{2k}$ in the second hidden layer $\boldsymbol{h}_{2}$ takes the
value $h_{2k}=\textrm{ReLU}(\boldsymbol{w}_{h_{2k}}^{T}\boldsymbol{h}_{1}+b_{h_{2k}})$,
and, in the end, 
\begin{align}
y_{\textrm{S\,(L)}} & =\textrm{Sigmoid}(\boldsymbol{w}_{y_{\textrm{S\,(L)}}}^{T}\boldsymbol{h}_{2}+b_{y_{\textrm{S\,(L)}}})\nonumber \\
 & \equiv(1+\exp(-(\boldsymbol{w}_{y_{\textrm{S\,(L)}}}^{T}\boldsymbol{h}_{2}+b_{y_{\textrm{S\,(L)}}})))^{-1}.
\end{align}
The sigmoid function is another nonlinear activation function \citep{Nielsen_Book_2015,Goodfellow_Book_2016},
whose return value is within $\left[0,1\right]$ and therefore can
be regarded as the ``confidence'' value. These two confidence values
$y_{\textrm{S}}$ and $y_{\textrm{L}}$ hold the meaning of how likely
any given configuration is a configuration of the solid and the liquid
phase, respectively, and the learnable parameter set $\left\{ \mathbf{w},\mathbf{b}\right\} \equiv\left\{ \boldsymbol{w}_{h_{1k}},\boldsymbol{w}_{h_{2k}},\boldsymbol{w}_{y_{\textrm{S\,(L)}}},b_{h_{1k}},b_{h_{2k}},b_{y_{\textrm{S\,(L)}}}\right\} $
completely determines the map that takes the system's configurations
as input and gives the confidence values $y_{\textrm{S}}$ and $y_{\textrm{L}}$
as output.

\begin{figure}
\noindent \begin{centering}
\includegraphics[width=3.3in]{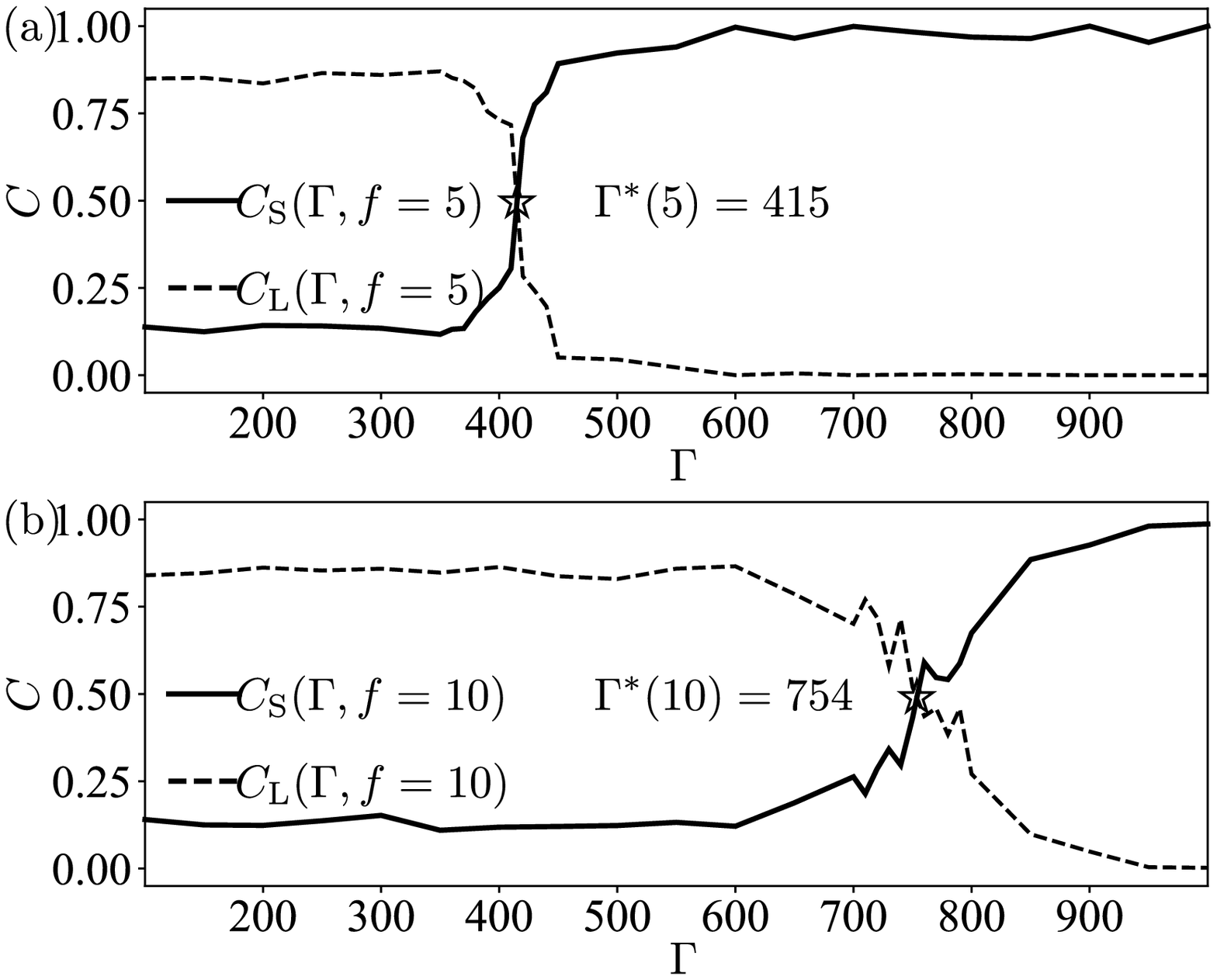}
\par\end{centering}
\caption{\label{fig:supervised_learning}Solid-liquid transition points predicted
by the data-driven criterion. The predicted transition points (marked
by stars) are (a) $\Gamma^{*}(5)=415$ and (b) $\Gamma^{*}(10)=754$,
which match very well with the predictions from the direct unsupervised
learning. Here, the self-propelled force strength $f=5$ is within
the parameter regime $f\in[0,8]$ where the supervised learning is
directly performed, while $f=10$ is way beyond $f\in[0,8]$. See
text for more details.}
\end{figure}

\section{TRAINING PROCESS OF NN: SUPERVISED LEARNING AND UNSUPERVISED LEARNING\label{sec:Training-process-of-NN}}

Here, we briefly outline how to train the NN, i.e., to optimize the
learnable parameter set $\left\{ \mathbf{w},\mathbf{b}\right\} $
with respect to the labels. For more thorough discussions on the machine
learning techniques involved in the training process, we refer the
reader to Refs.~\citep{Nielsen_Book_2015,Goodfellow_Book_2016}.

To train the NN, one shall first label the data, where a label $\tilde{\boldsymbol{y}}=(\tilde{y}_{\textrm{S}},\tilde{y}_{\textrm{L}})^{T}$
is a suggestion of the expected $y_{\textrm{S}}$ and $y_{\textrm{L}}$.
For instance, while labeling an image $I$ as ``liquid'', we suggest
that $I$ is $100\%$ likely to be a configuration of the liquid phase,
namely, $\tilde{\boldsymbol{y}}=(0,1)^{T}$. In both the supervised
learning process and the unsupervised learning process in this work,
every sample of data used for training has a label. The major difference
between these two learning process is whether the labels are physical.
More specifically, labels in the supervised learning process are always
expected to be physically correct, while labels in the unsupervised
learning process are not always the case. After the labeling, the
error of the confidence values compared to $\tilde{\boldsymbol{y}}$
can be quantified by the cross-entropy cost function $S=-(\tilde{y}_{\textrm{S}}\ln y_{\textrm{S}}+\tilde{y}_{\textrm{L}}\ln y_{\textrm{L}})$
\citep{Nielsen_Book_2015,Goodfellow_Book_2016}. Whenever the NN is
``trained'' in this work, $\left\{ \mathbf{w},\mathbf{b}\right\} $
is optimized by minimizing the cost function $S$ traversing the training
samples for $10$ epochs using the Adam method \citep{Kingma_arXiv_2014}
with the learning rate $1\times10^{-3}$.

\begin{figure}
\noindent \begin{centering}
\includegraphics[width=3.3in]{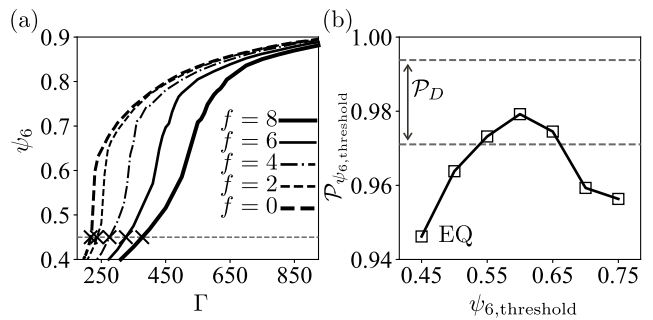}
\par\end{centering}
\caption{\label{fig:NEQ_Psi6_threshold}(a) $\Gamma$-dependence of the global
bond-orientational order parameter $\psi_{6}$ of the system at different
self-propelled force strengths $f=0,2,4,6,8$. The horizontal dashed
line corresponds to the equilibrium threshold $\psi_{6\textrm{,threshold}}^{\textrm{EQ}}=0.45$.
The crossings marked by ``$\times$'' of the horizontal dashed line
with the $D(\Gamma)$ curves at different $f$ predict the corresponding
transition points $\Gamma_{\psi_{6}}^{*}(f)$. (b) Threshold value
$\psi_{6\textrm{,threshold}}$ dependence of the criterial classification
accuracy $\mathcal{P}_{\psi_{6\textrm{,threshold}}}$. The leftmost
point corresponds to $\psi_{6\textrm{,threshold}}=\psi_{6\textrm{,threshold}}^{\textrm{EQ}}=0.45$,
and the point with the highest $\mathcal{P}_{\psi_{6\textrm{,threshold}}}$
corresponds to $\psi_{6\textrm{,threshold}}=0.6$ which is a general
threshold value for $\psi_{6}$ in the NEQ scenario. For comparison,
the highest and the lowest criterial classification accuracy associated
with the long-time diffusion coefficient $D$, i.e., $\mathcal{P}_{D_{\mathrm{threshold}}^{\mathrm{NEQ}}}$
and $\mathcal{P}_{D_{\mathrm{threshold}}^{\mathrm{EQ}}}$, are shown
by the upper and the lower horizontal dashed lines, respectively.
See text for more details.}
\end{figure}

\section{TRANSITION POINTS PREDICTED BY THE DATA-DRIVEN CRITERION\label{sec:Transition-points-predicted}}

The data-driven criterion which captures the generic feature that
distinguishes the solid from the liquid phase is expected to give
equal confidence values $y_{\textrm{S}}$ and $y_{\textrm{L}}$ at
the solid-liquid transition point, since the system can be either
in the solid phase or the liquid phase. Therefore, after the supervised
learning process with all the properly labeled data at $f=0,2,4,6,8$
in three independent data sets ``A,'' ``B,'' and ``C,'' the
solid-liquid transition point for a fixed self-propelled force strength
$f$ can be predicted by calculating the average classification confidence
values \citep{Melko_Nat_Phys_2017}, whose explicit form reads $C_{\textrm{S\,(L)}}(\Gamma,f)=\sum_{n=1}^{\mathcal{N}_{\Gamma,f}}y_{\textrm{S\,(L)}}(I_{\Gamma,f;n})\slash\mathcal{N}_{\Gamma,f}$,
where $I_{\Gamma,f;n}$ is the $n$th testing sample with its corresponding
effective interaction strength being $\Gamma$ and $\mathcal{N}_{\Gamma,f}$
is the total number of these samples. Here, the intersection points
of the $\Gamma$-dependence curves of $C_{\textrm{S}}$ and $C_{\textrm{L}}$
in Fig.~\ref{fig:supervised_learning} indicate $\Gamma^{*}(5)=415$
and $\Gamma^{*}(10)=754$, respectively, which match very well with
the predictions from the direct unsupervised learning.

Moreover, as a piece of technical information, we provide here more
details concerning the data set used in this work. For each different
value of $(\Gamma,f)$, there are $7.2\times10^{3}$ samples generated
in the direct numerical simulations of the dynamical equation (\ref{eq:Brownian_Yukawa}).
The spacing for different $f$ is kept the same with $f=0,2,4,5,6,8,10$.
In the expected deep solid and deep liquid regions, the spacing $\Delta\Gamma$
between different $\Gamma$ assumes a relatively larger value $\Delta\Gamma=50$,
with $\Gamma=50,100,\ldots,1000$. While in the rest of the parameter
region $\Delta\Gamma=10$, with $\Gamma=210,220,\ldots,290$ for $f=0$
and $f=2$; $\Gamma=310,320,\ldots,390$ for $f=4$; $\Gamma=360,370,\ldots,440$
for $f=5$; $\Gamma=410,420,\ldots,490$ for $f=6$; $\Gamma=560,570,\ldots,640$
for $f=8$; $\Gamma=710,720,\ldots,790$ for $f=10$.

\begin{figure}
\noindent \begin{centering}
\includegraphics[width=3.3in]{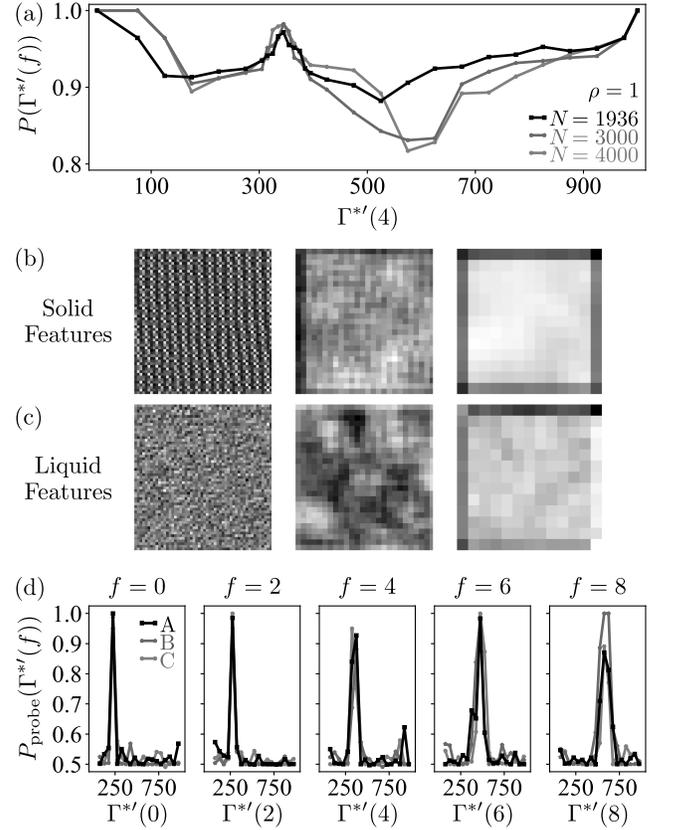}
\par\end{centering}
\caption{\label{fig:CNN}(a) Solid-liquid transition point at self-propelled
force strength $f=4$ identified by the CNN trained via the ``learning
by confusion'' unsupervised learning, with a larger data set that
is organized differently from the data set that corresponds to Fig.~\ref{fig:unsupervised_learning}(b).
For different particle number $N$ at constant global density $\rho=1$,
the CNN's classification accuracy $P(\Gamma^{*\prime}(f))$ reaches
the same nontrivial maximum $\Gamma^{*}(4)=345\pm5$ as the one obtained
by the fully connected NN. This suggests the finite-size effects on
the phase transition points are quite small at the system size mainly
employed in this work (i.e., $N=1936$ and $\rho=1$). (b) Typical
features implied in the weight map of the well-trained CNN\textquoteright s
first three convolutional layers concerning a sample of the solid
phase with $\Gamma=1000$ at $f=4$, where the lighter color corresponds
to the higher level of excitation. The weight map of the first convolutional
layer (left panel) manifests a clear periodic structure, and for deeper
layers (middle and right panels), it assumes a relatively homogeneous
structure. (c) Analogous data to (b), but concerning a sample of the
liquid phase with $\Gamma=50$ at $f=4$. The weight maps manifest
a relatively disordered structure. (d) Solid-liquid transition points
identified by the CNN trained via another machine learning approach
that can detect the possible intermediate phases. Via calculating
the signal $P_{\textrm{probe}}(\Gamma^{*\prime}(f))$ to probe the
phase transition in the parameter region of interests, only one phase
transition can be detected at each $f$ within a finite resolution
$\Delta\Gamma=25$, indicating that this finite-size system assumes
only two phases, which is consistent with previous investigations.
And the predicted transition points agree well with the results in
Fig.~\ref{fig:unsupervised_learning}(b). See text for more details.}
\end{figure}

\section{CRITERIAL CLASSIFICATION ACCURACY AND THE NEQ THRESHOLD VALUE\label{sec:Criterial-classification-accuracy}}

For a given criterion $\mathcal{C}$, the transition point $\Gamma_{\mathcal{C}}^{*}(f_{i})$
at the self-propelled force strength $f_{i}$ is determined, and it
corresponds to a binary classification rule associated with the proposed
critical value $\Gamma^{*\prime}(f_{i})=\Gamma_{\mathcal{C}}^{*}(f_{i})$.
Therefore, after the unsupervised learning process, the criterial
classification accuracy $\mathcal{P}_{\mathcal{C}}=\sum_{i=1}^{N_{f}}P(\Gamma_{\mathcal{C}}^{*}(f_{i}))\slash N_{f}$
can be directly calculated according to the $\Gamma^{*\prime}(f_{i})$
dependence curves of the classification accuracy $P(\Gamma^{*\prime}(f_{i}))$.
Concerning the empirical criterion associated with the long-time diffusion
coefficient $D$, we monitor the criterial classification accuracy
$\mathcal{P}_{D_{\mathrm{threshold}}}=\sum_{f=0,2,4,6,8}P(\Gamma_{D_{\mathrm{threshold}}}^{*}(f))/5$
by utilizing all the available data at $f=0,2,4,6,8$ in three independent
data sets ``A'', ``B'', and ``C'', and indeed find an optimal
choice of a smaller NEQ threshold $D_{\textrm{threshold}}^{\textrm{NEQ}}=0.025$.
However, for the empirical criterion associated with the global bond-orientational
order parameter $\psi_{6}$, it assumes relatively poor performance
in general. As we can see from Fig.~\ref{fig:NEQ_Psi6_threshold}(b),
for this empirical criterion, the highest criterial classification
accuracy $\mathcal{P}_{\psi_{6\textrm{,threshold}}}$ with a general
threshold value $\psi_{6\textrm{,threshold}}=0.6$ for $\psi_{6}$
is just roughly around $\mathcal{P}_{D_{\mathrm{threshold}}^{\mathrm{EQ}}}$,
and much lower than $\mathcal{P}_{D_{\mathrm{threshold}}^{\mathrm{NEQ}}}$
{[}cf.~Figs.~\ref{fig:NEQ_D_threshold}(b) and \ref{fig:NEQ_Psi6_threshold}(b){]}.

\section{RESULTS BY EMPLOYING CONVOLUTIONAL NN\label{sec:Convolutional-neural-network}}

In the main text, we have established the data-driven criterion and
the data-driven evaluation function by applying the hybrid machine
learning approach realized by a fully connected NN. To address a few
relevant technical aspects, here we present the physical results obtained
by the same approach employing a NN with a different architecture,
namely, a convolutional NN (CNN) called ``AlexNet'' \citep{Krizhevsky_Adv_Neural_Inf_Process_Syst_2012}
(with regularization implemented via ``dropout''). As we shall see
in the following, these results are the same as the ones obtained
via the fully connected NN.

Here, we take the case with $f=4$ for instance. We perform the ``learning
by confusion'' unsupervised learning using the CNN. Moreover, besides
the NN with a different architecture employed to obtain the results
shown in Fig.~\ref{fig:CNN}(a), the data set that corresponds to
Fig.~\ref{fig:CNN}(a) is also organized differently from the one
that corresponds to Fig.~\ref{fig:unsupervised_learning}(b). For
the former, all the samples generated by the same simulation are either
used in the training, the validation, or the test \citep{Nielsen_Book_2015,Goodfellow_Book_2016},
while for the latter, the samples generated by the same simulation
are divided into three parts for the training, the validation, and
the test, respectively. In addition, for the former, the total number
of samples in the data set is twice as large as the one for the later.
As we can see from the black curve in Fig.~\ref{fig:CNN}(a), comparing
with the middle column of Fig.~\ref{fig:unsupervised_learning}(b),
the CNN's classification accuracy $P(\Gamma^{*\prime}(f))$ reaches
the same nontrivial maximum as the one obtained by the fully connected
NN. This manifests that the physical results obtained by the machine
learning approach in this work are robust against these technical
differences in its realization. 

Moreover, we further study the influence of finite-size effects on
the phase transition points. More specifically, we perform numerical
calculations at different system sizes with the total particle number
$N=1936,3000,4000$ and the global density $\rho=1$ keeping at constant.
As we can see from Fig.~\ref{fig:CNN}(a), the predicted solid-liquid
transition points (at $f=4$) of the system with larger sizes are
numerically the same as the one of the system with $N=1936$ and $\rho=1$,
i.e., $\Gamma^{*}(4)=345\pm5$. This suggests that finite-size effects
on the phase transition points are quite small at the system size
($N=1936$ and $\rho=1$) mainly employed in this work.

Finally, let us take the chance of utilizing the easily-accessed weight
map \citep{Nielsen_Book_2015,Goodfellow_Book_2016} of the CNN to
visualize the feature used by the NN to perform the classification.
In Figs.~\ref{fig:CNN}(b) and \ref{fig:CNN}(c), we show typical
features implied in the weight map \citep{Nielsen_Book_2015,Goodfellow_Book_2016}
of the well-trained CNN\textquoteright s first three convolutional
layers, where the lighter color corresponds to the higher level of
excitation \citep{Goodfellow_IEEE_T_PAMI_2013}. The weight map of
the first convolutional layer of the CNN manifests a clear periodic
structure for the samples of the solid phase with $\Gamma=1000$ at
$f=4$ {[}cf.~the left panel of Fig.~\ref{fig:CNN}(b){]}, while
a disordered structure for the ones of the liquid phase with $\Gamma=50$
at $f=4$ {[}cf.~the left panel of Fig.~\ref{fig:CNN}(c){]}. For
deeper layers, the corresponding weight map assumes a relatively homogeneous
structure for the solid phase {[}cf.~middle and right panels of Fig.~\ref{fig:CNN}(b){]},
and a relatively disordered structure for the liquid phase {[}cf.~middle
and right panels of Fig.~\ref{fig:CNN}(c){]}. From the general network
structure of the CNN \citep{Nielsen_Book_2015,Goodfellow_Book_2016},
the behavior of the weight maps seems to suggest that the feature
used by the CNN to classify samples is closely related to the spatial
Fourier transformation of the system\textquoteright s density distribution.

\section{INTERMEDIATE PHASE DETECTION\label{sec:scanning_probe}}

In the main text, the unsupervised part of the hybrid machine learning
approach is limited to the cases where maximally two possible phases
exist in the parameter regimes of interests. The major reason for
choosing this approach comes from the fact that previous investigations
have not resolved the possible (hexatic) intermediate phase in the
system with a finite size \citep{Lowen_PRL_2012}.

To further check this point, here we use another machine learning
approach developed recently \citep{Guo_arXiv_2021b} that can detect
the possible intermediate phases. In this new approach \citep{Guo_arXiv_2021b},
the signal to probe the phase transition, denoted as $P_{\textrm{probe}}(\Gamma^{*\prime}(f))$,
is calculated for each narrow parameter interval $[\Gamma^{*\prime}(f)-\Delta\Gamma,\Gamma^{*\prime}(f)+\Delta\Gamma]$,
with $\Gamma^{*\prime}(f)$ being the suspected phase transition point
and $\Delta\Gamma$ being the resolution of the probe. Here, $P_{\textrm{probe}}(\Gamma^{*\prime}(f))$
is nothing but the classification accuracy of the NN concerning the
data associated with the two boundaries of the corresponding interval,
which indicates how likely there is a phase transition in the interval.
As we can see from Fig.~\ref{fig:CNN}(d), via calculating $P_{\textrm{probe}}(\Gamma^{*\prime}(f))$
in the parameter region of interests, only one phase transition can
be detected at each $f$ within a finite resolution $\Delta\Gamma=25$,
indicating that this finite-size system assumes only two phases, which
is consistent with previous investigations \citep{Lowen_PRL_2012}.
We can also notice that the predicted transition points agree well
with the results in Fig.~\ref{fig:unsupervised_learning}(b) in the
main text.

\end{document}